\documentclass[aps,prc,amsmath,two column,amssymb,showpacs]{revtex4}
\usepackage{txfonts}
\usepackage{mathbbold}
\usepackage{epsfig,bm,dcolumn}
\usepackage{graphicx}
\usepackage{color}

\begin{document}

\title{Unusual Zeeman-field effects in two-dimensional spin-orbit-coupled Fermi superfluids}

\author{Lianyi He$^1$}
\author{Xu-Guang Huang$^{1,2}$}
\affiliation{$^1$ Frankfurt Institute for Advanced Studies and Institute for Theoretical Physics,
J. W. Goethe University, 60438 Frankfurt am Main, Germany\\
$^2$ Center for Exploration of Energy and Matter and Physics Department,
Indiana University, Bloomington, IN 47408, USA}

\date{\today}

\begin{abstract}
We investigate the Zeeman field effects on the bulk superfluid properties and the collective modes in two-dimensional (2D) attractive atomic Fermi gases with Rashba-type spin-orbit coupling. In the presence of a large spin-orbit coupling, the system undergoes a quantum phase transition to a topological superfluid state at a critical Zeeman field. We show that the nonanalyticities of the thermodynamic functions as well as other physical quantities at the quantum phase transition originate from the infrared singularities caused by the gapless fermionic spectrum. The same argument applies also to the BCS-BEC evolution in 2D fermionic superfluids with $p$- or $d$-wave pairing. The superfluid density $n_s$ and the velocity of the Goldstone sound mode $c_s$ behave oppositely in the normal and the topological superfluid phases: they are suppressed by the Zeeman field in the normal superfluid phase, but get enhanced in the topological superfluid phase. The velocity of the Goldstone sound mode also shows nonanalyticity at the quantum phase transition. For large Zeeman field, we find $n_s\rightarrow n$ and $c_s\rightarrow \upsilon_{\rm F}$, where $n$ is the total fermion density and $\upsilon_{\rm F}$ is the Fermi velocity of noninteracting system. The unusual behavior of the superfluid density and the collective modes can be understood by the fact that the spin-orbit-coupled superfluid state at large Zeeman field can be mapped to the $p_x+ip_y$ superfluid state of spinless fermions.
\end{abstract}

\pacs{03.75.Ss, 05.30.Fk, 67.85.Lm, 74.20.Fg}

\maketitle

The Zeeman field (ZF, denoted by $h$) effects on BCS superconductivity have been a longstanding problem for several decades~\cite{LOFFreview}. At weak coupling, the BCS state undergoes a first-order phase transition to the normal state at $h_{\rm CC}=0.707\Delta_0$~\cite{CClimit} where $\Delta_0$ is the pairing gap at $h=0$. Further studies showed that the inhomogeneous Fulde-Ferrell-Larkin-Ovchinnikov (FFLO) state~\cite{FFLO} survives in a narrow window between $h_{\rm CC}$ and $h_{\rm FFLO}=0.754\Delta_0$. The ZF effects on the fermionic superfluidity in the whole BCS-BEC crossover~\cite{BCSBEC} regime have been experimentally studied in recent years~\cite{Imexp}. Two-component atomic Fermi gases with population imbalance were realized to simulate the ZF effects. Around the Feshbach resonance, the phase separation between the superfluid and the normal phases has been observed in accordance with the first-order phase transition. Despite the rich phase structure in the BCS-BEC crossover~\cite{Imth}, the superfluidity is completely destroyed at large enough ZF.

Recent progress on synthetic spin-orbit coupling (SOC) for neutral atoms~\cite{SOCth,SOCB,SOCF} provides new ways to study SOC effects on fermionic superfluidity~\cite{SOC-BCSBEC}. Previous studies of two-dimensional (2D) solid-state systems showed that the SOC induces spin-triplet pairing, even though the attractive interaction is $s$ wave~\cite{Gorkov}. By applying a large ZF, the 2D system undergoes a topological phase transition to a topological superconducting state, where the non-Abelian topological order and Majorana fermionic modes can be realized~\cite{TSC01}. However, the properties of the bulk phase transition and the collective modes are less understood for such systems.

In this paper, we study the bulk phase transition and the collective modes in 2D atomic Fermi gases with combined SOC and ZF effects. The main results can be summarized as follows:
(i) The bulk phase transition originates from the infrared singularities caused by the gapless fermionic spectrum. The analyticity of any physical quantity across the phase transition can be determined by analyzing the infrared behavior of the momentum integrals.
For the present system, we find that the quantum phase transition is of third order.
(ii) The superfluid density $n_s$ and the velocity of the Goldstone sound mode $c_s$ behave oppositely in the normal and the topological superfluid phases. They are suppressed by the ZF in the normal superfluid phase but turn to increase with the ZF in the topological superfluid phase. The sound velocity $c_s$ also shows nonanalyticity across the phase transition.
(iii) For very large ZF ($h\rightarrow\infty$), we obtain analytically $n_s\rightarrow n$ and $c_s\rightarrow \upsilon_{\rm F}$, where $n$ is the total fermion density and $\upsilon_{\rm F}$ is the Fermi velocity of noninteracting systems. We show that the unusual behavior of the superfluid density and the collective modes is manifest in the fact that the spin-orbit-coupled superfluid state at large ZF can be mapped to the $p_x+ip_y$ superfluid state of spinless fermions.

\emph{Model and effective potential -- } The many-body Hamiltonian for the 2D Fermi system we considered can be written as
$H =H_{\rm s}+H_{\rm int}$, where
\begin{eqnarray}
&&H_{\rm s}=\int d^2 {\bf r}\psi^{\dagger}({\bf r})
\left(\frac{\hat{\bf p}^2}{2M}-\mu+{\cal H}_{\rm SO}+{\cal H}_Z\right) \psi({\bf r}),\nonumber\\
&&H_{\rm int}=-U\int d^2 {\bf r}^{\phantom{\dag}}\psi^\dagger_{\uparrow}({\bf r})
\psi^\dagger_{\downarrow}({\bf r})\psi^{\phantom{\dag}}_{\downarrow}({\bf r})\psi^{\phantom{\dag}}_{\uparrow}({\bf r}).
\end{eqnarray}
Here, $\psi({\bf r})= [\psi_\uparrow({\bf r}), \psi_\downarrow({\bf r})]^{\rm T}$ represents the two-component fermion fields, $\hat{\bf p}
=\hat{p}_x{\bf e}_x+\hat{p}_y{\bf e}_y$ is the 2D momentum operator with $\hat{p}_i=-i\hbar\partial_i$, $\mbox{\boldmath{$\sigma$}}=\sigma_x
{\bf e}_x+\sigma_y{\bf e}_y$ with $\sigma_i$ being the Pauli matrices, and $\mu$ is the chemical potential. The contact coupling $U>0$ denotes the attractive $s$-wave interaction between unlike spins. The ZF term reads ${\cal H}_Z=-h\sigma_z$ and the spin-dependent term ${\cal H}_{\rm SO}
=\lambda\mbox{\boldmath{$\sigma$}}\cdot\hat{\bf p}$ is the 2D SOC~\cite{RSOC}. We set $h>0$ and $\lambda>0$ without loss of generality. In the following we use the units $\hbar=k_{\rm B}=M=1$.

In the imaginary-time functional integral formalism (temperature $T=1/\beta$), the partition function of the system is
${\cal Z} = \int\mathcal{ D}\psi\mathcal{D}\psi^\dagger\exp\left\{-{\cal S}[\psi,\psi^\dagger]\right\}$ with the action
${\cal S}[\psi,\psi^\dagger]=\int_0^\beta d\tau\left[\int d^2{\bf r}\psi^\dagger\partial_\tau \psi+H(\psi,\psi^\dagger)\right]$. Introducing the
pair field $\Phi(x) = -U\psi_\downarrow(x)\psi_\uparrow(x)$~$[x=(\tau,{\bf r})]$ and integrating out the fermionic degrees of freedom, we obtain
$\mathcal {Z}=\int\mathcal{D} \Phi \mathcal{D} \Phi^{\dagger} \exp \big\{- {\cal S}_{\rm{eff}}[\Phi, \Phi^{\dagger}]\big\}$,
where the effective action is given by
\begin{eqnarray}
{\cal S}_{\rm{eff}}[\Phi, \Phi^{\dagger}] = \frac{1}{U}\int dx|\Phi(x)|^{2} - \frac{1}{2}{\rm{Trln}} [{\bf G}^{-1}(x,x^\prime)].
\end{eqnarray}
In the Nambu-Gor'kov representation, the inverse single-particle Green's function reads
\begin{eqnarray}
{\bf G}^{-1}(x,x^\prime)=\left(\begin{array}{cc}{\bf G}_+^{-1}(x)&\Phi(x)\\
\Phi^\dagger(x)& {\bf G}_-^{-1}(x)\end{array}\right)\delta(x-x^\prime),
\end{eqnarray}
where ${\bf G}_\pm^{-1}(x)=-\partial_{\tau}+h\sigma_z \mp(\hat{\bf p}^2/2+\lambda\mbox{\boldmath{$\sigma$}}\cdot\hat{\bf p}-\mu)$.

In the superfluid state, the pairing field $\Phi(x)$ acquires a nonzero expectation value $\langle\Phi(x)\rangle=\Delta$ which we set to be real without loss of generality. By separating the pairing field as $\Phi(x)=\Delta+\phi(x)$, the effective action
${\cal S}_{\text{eff}}[\Phi,\Phi^\dagger]$ can be expanded in powers of the complex fluctuation field $\phi(x)$. We have
\begin{equation}
{\cal S}_{\text{eff}}[\Phi,\Phi^\dagger]={\cal S}_{\text{eff}}^{(0)}(\Delta)+{\cal S}_{\text{eff}}^{(2)}[\phi,\phi^\dagger]+\cdots,
\end{equation}
where ${\cal S}_{\text{eff}}^{(0)}(\Delta)\equiv{\cal S}_{\rm eff}[\Delta,\Delta]$ is the saddle-point or mean-field effective action with the pair potential $\Delta$ determined by the saddle point condition $\partial{\cal S}_{\text{eff}}^{(0)}/\partial\Delta=0$. The collective modes are determined by the Gaussian-fluctuation part ${\cal S}_{\text{eff}}^{(2)}[\phi,\phi^\dagger]$.

\emph{Infrared singularity and bulk phase transition -- } The single-particle excitation spectra can be read from the pole of the fermion Green's function ${\cal G}(K)$, which is obtained from ${\bf G}$ by the replacement $\Phi\rightarrow\Delta$. Here, $K=(i\omega_n,{\bf k})$ with $\omega_n$ being the fermion Matsubara frequency. Working out the explicit form of ${\cal G}(K)$, we obtain the quasiparticle dispersion $\pm E_{\bf k}^\alpha$ ($\alpha=\pm$), where $E_{\bf k}^\alpha$ is given by
\begin{eqnarray}
E_{\bf k}^\alpha=\sqrt{E_{\bf k}^2+\eta_{\bf k}^2+2\alpha\zeta_{\bf k}}.
\end{eqnarray}
Here we have defined $E_{\bf k}=(\xi_{\bf k}^2+\Delta^2)^{1/2}$, $\eta_{\bf k}=(\lambda^2{\bf k}^2+h^2)^{1/2}$, and
$\zeta_{\bf k}=(\xi_{\bf k}^2\eta_{\bf k}^2+h^2\Delta^2)^{1/2}$ with $\xi_{\bf k}={\bf k}^2/2-\mu$. From the identity
$(E_{\bf k}^+)^2(E_{\bf k}^-)^2=(E_{\bf k}^2-\eta_{\bf k}^2)^2+4\lambda^2{\bf k}^2\Delta^2$, we find that the fermionic excitations are fully gapped for $\Delta\neq0$ except for the case that the condition ${\cal C}_0=\mu^2+\Delta^2-h^2=0$ is satisfied. For ${\cal C}_0=0$, the lower branch
$E_{\bf k}^-$ has a linear dispersion near ${\bf k}=0$; that is, $E_{\bf k}^-=\upsilon_c|{\bf k}|+O(|{\bf k}|^2)$, where the velocity $\upsilon_c=\lambda \Delta/h$.

The gapless fermionic spectrum causes nonanalyticities of some physical quantities at the critical point ${\cal C}_0=0$. To be specific, we consider the thermodynamic potential $\Omega(\mu,h)\equiv\Omega(\mu,h,\Delta(\mu,h))$ at zero temperature, where
\begin{eqnarray}
\Omega(\mu,h,\Delta)=\sum_{\bf k}\left(\frac{\Delta^2}{{\bf k}^2+\epsilon_{\rm B}}-\frac{E_{\bf k}^++E_{\bf k}^-}{2}+\xi_{\bf k}\right).
\end{eqnarray}
Here we have used the usual regularization $U^{-1}=\sum_{\bf k}({\bf k}^2+\epsilon_{\rm B})^{-1}$ for 2D systems~\cite{BCSBEC2D} with
$\epsilon_{\rm B}$ being the binding energy of the two-body bound state in the absence of SOC. To obtain the thermodynamic potential $\Omega(\mu,h)$, the pair potential $\Delta(\mu,h)$, which is regarded as an implicit function of $\mu$ and $h$, should be determined by the gap equation
$\partial \Omega(\mu,h,\Delta)/\partial\Delta=0$.

To study the analyticity of the thermodynamic potential or its derivatives with respect to $\mu$ and $h$, we consider the following susceptibilities:
\begin{equation}
\chi_{\mu\mu}=-\frac{\partial^2\Omega(\mu,h)}{\partial \mu^2},\ \ \ \ \chi_{hh}=-\frac{\partial^2\Omega(\mu,h)}{\partial h^2},
\end{equation}
which are related to the isothermal compressibility and the spin susceptibility, respectively. To obtain their explicit expressions, we need the derivatives $\partial\Delta(\mu,h)/\partial\mu$ and $\partial\Delta(\mu,h)/\partial h$. They can be obtained from the gap equation
$\partial \Omega(\mu,h,\Delta)/\partial\Delta=0$. Finally, the two susceptibilities can be evaluated as
\begin{eqnarray}
\chi_{\mu\mu}&=&\frac{\partial n(\mu,h,\Delta)}{\partial \mu}+\frac{1}{A}\left(\frac{\partial n(\mu,h,\Delta)}{\partial \Delta}\right)^2,\nonumber\\
\chi_{hh}&=&\frac{\partial m(\mu,h,\Delta)}{\partial h}+\frac{1}{A}\left(\frac{\partial m(\mu,h,\Delta)}{\partial \Delta}\right)^2.
\end{eqnarray}
Here $A=\partial^2\Omega(\mu,h,\Delta)/\partial \Delta^2$, $n=-\partial \Omega(\mu,h,\Delta)/\partial\mu$ is the total density, and
$m=-\partial \Omega(\mu,h,\Delta)/\partial h$ is the spin polarization.

We find that the expressions of $\chi_{\mu\mu}$ and $\chi_{hh}$ contain some momentum integrals of the following type:
\begin{eqnarray}
{\cal I}_{ij}\sim \int_0^\infty kdk \frac{{\cal Q}_i{\cal Q}_j}{(E_{\bf k}^-)^3}g(k),
\end{eqnarray}
where ${\cal Q}_1=1-h^2/\zeta_{\bf k}$, ${\cal Q}_2=1-\eta_{\bf k}^2/\zeta_{\bf k}$, and ${\cal Q}_3=1-E_{\bf k}^2/\zeta_{\bf k}$. The function $g(k)$ approaches some nonzero constant for $k\rightarrow0$. At ${\cal C}_0=0$, the integrals ${\cal I}_{ij}$ are infrared safe since the quantities ${\cal Q}_i$ go as $k^2$ for $k\rightarrow 0$. Therefore, $\chi_{\mu\mu}$ and $\chi_{hh}$ are continuous across the phase transition. However, the $l$-th derivatives of the susceptibilities with respect to $\mu$ or $h$ contain momentum integrals whose infrared behavior goes as
\begin{eqnarray}
\int_0^\epsilon kdk \frac{k^{4-2l}}{k^3}=\int_0^\epsilon dk k^{2-2l}.
\end{eqnarray}
For $l=2$, the infrared divergence shows up. Therefore, the fourth derivative of $\Omega(\mu,h)$ is divergent at the phase transition. Then the third derivative is discontinuous and hence the susceptibilities show nonanalyticities. Based on these observations, we conclude that the quantum phase transition at ${\cal C}_0=0$ is of third order~\cite{note01}.

\begin{figure*}[!htb]
\begin{center}
\includegraphics[width=7cm]{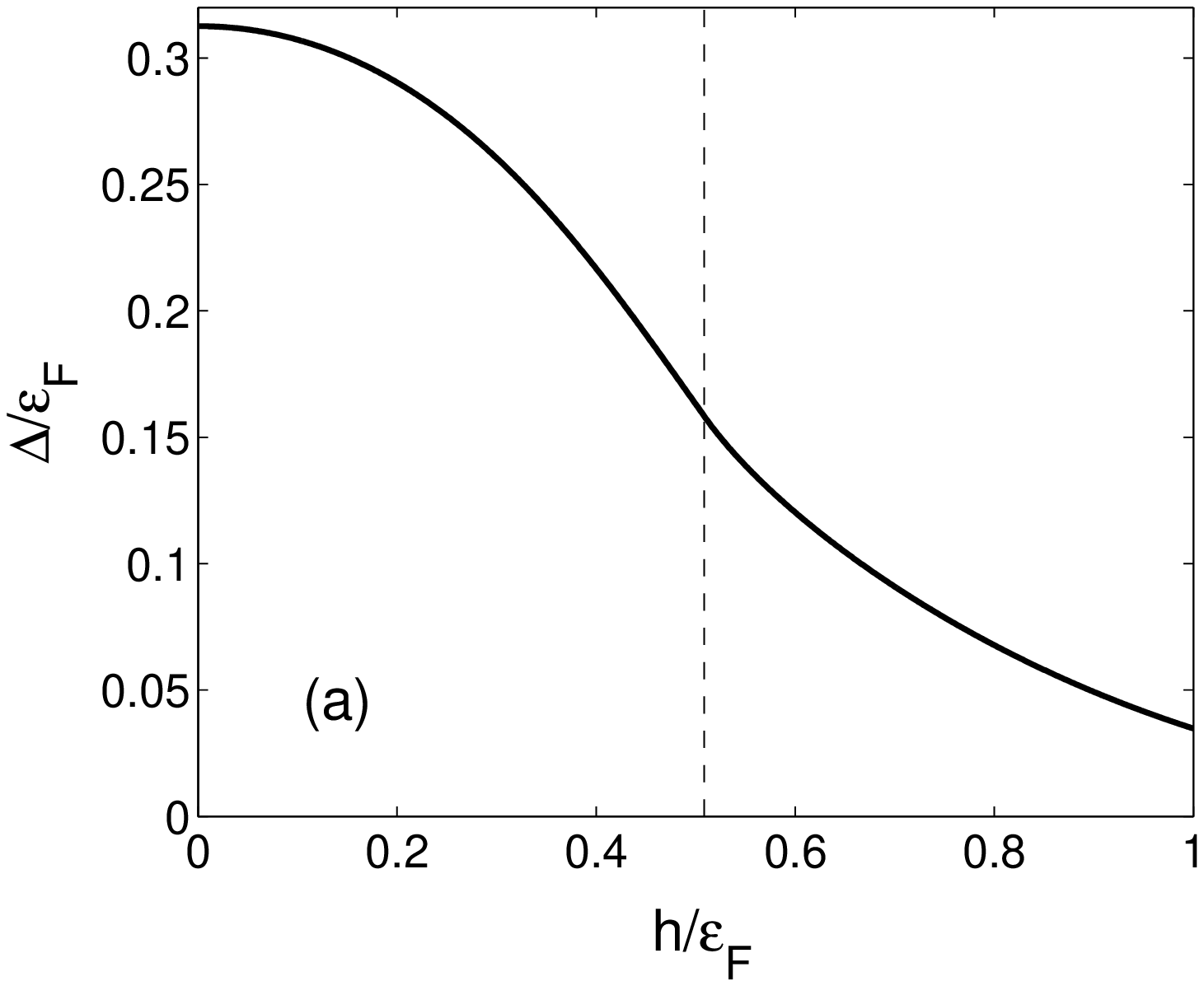}
\includegraphics[width=7cm]{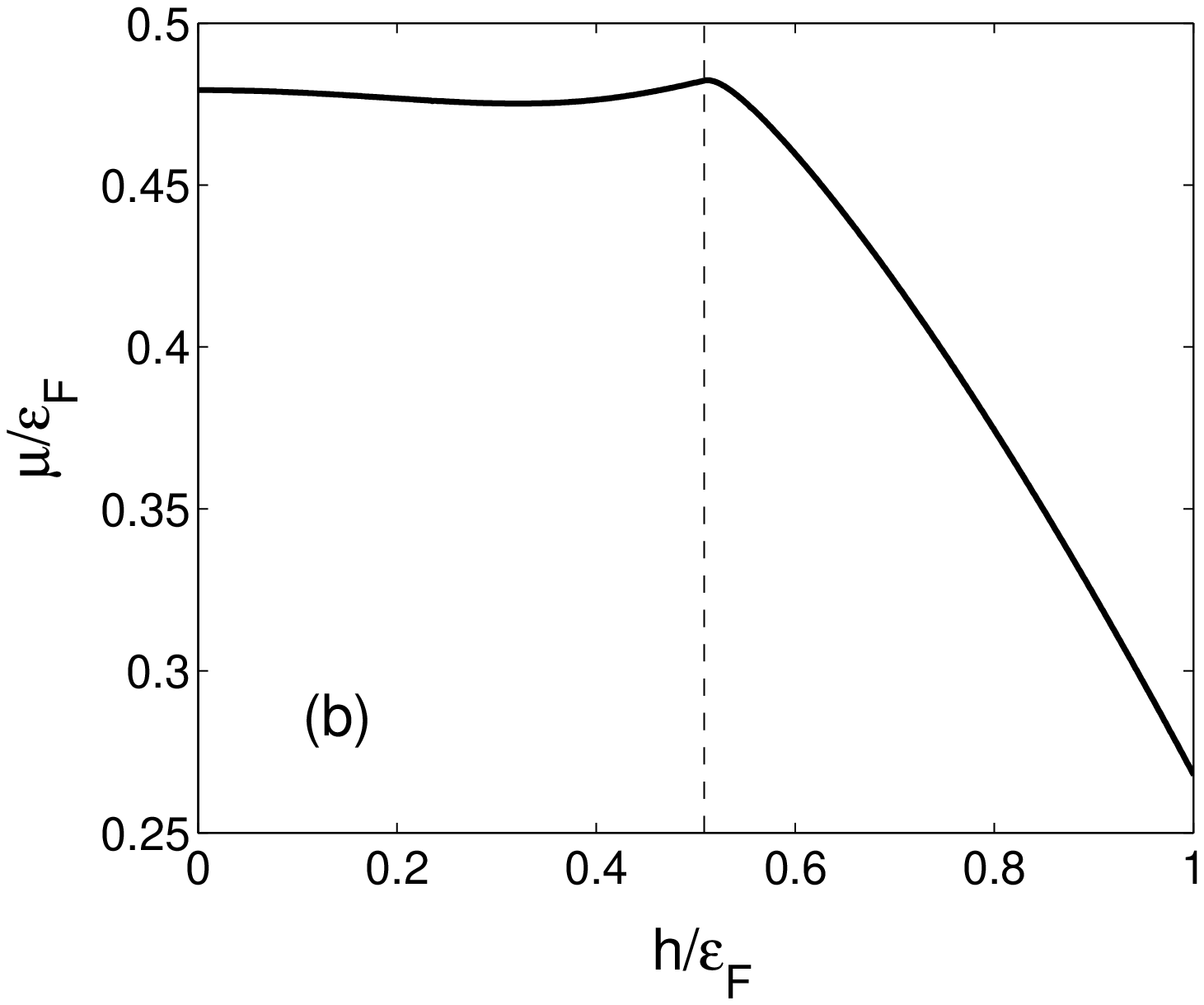}
\includegraphics[width=7cm]{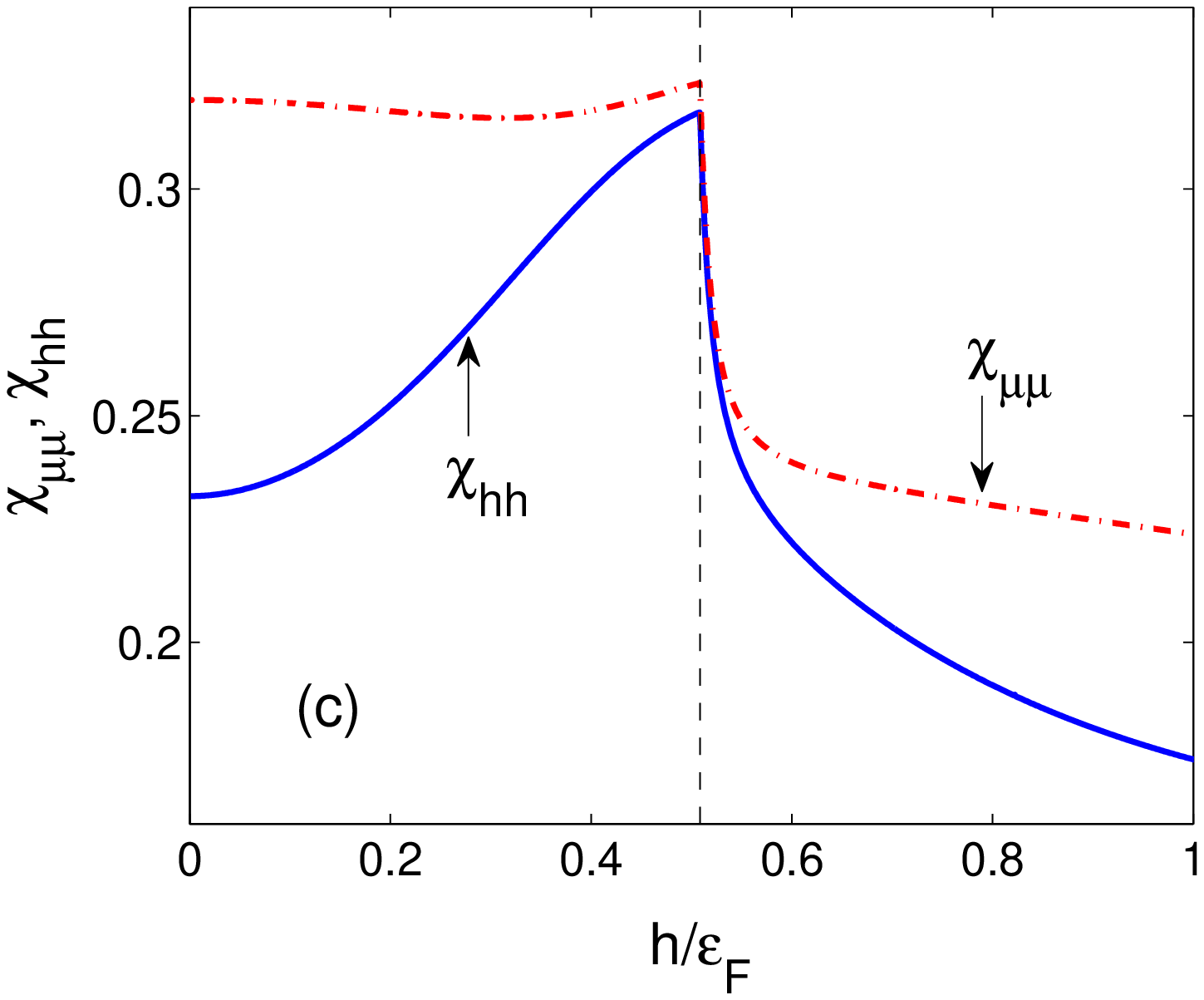}
\includegraphics[width=7cm]{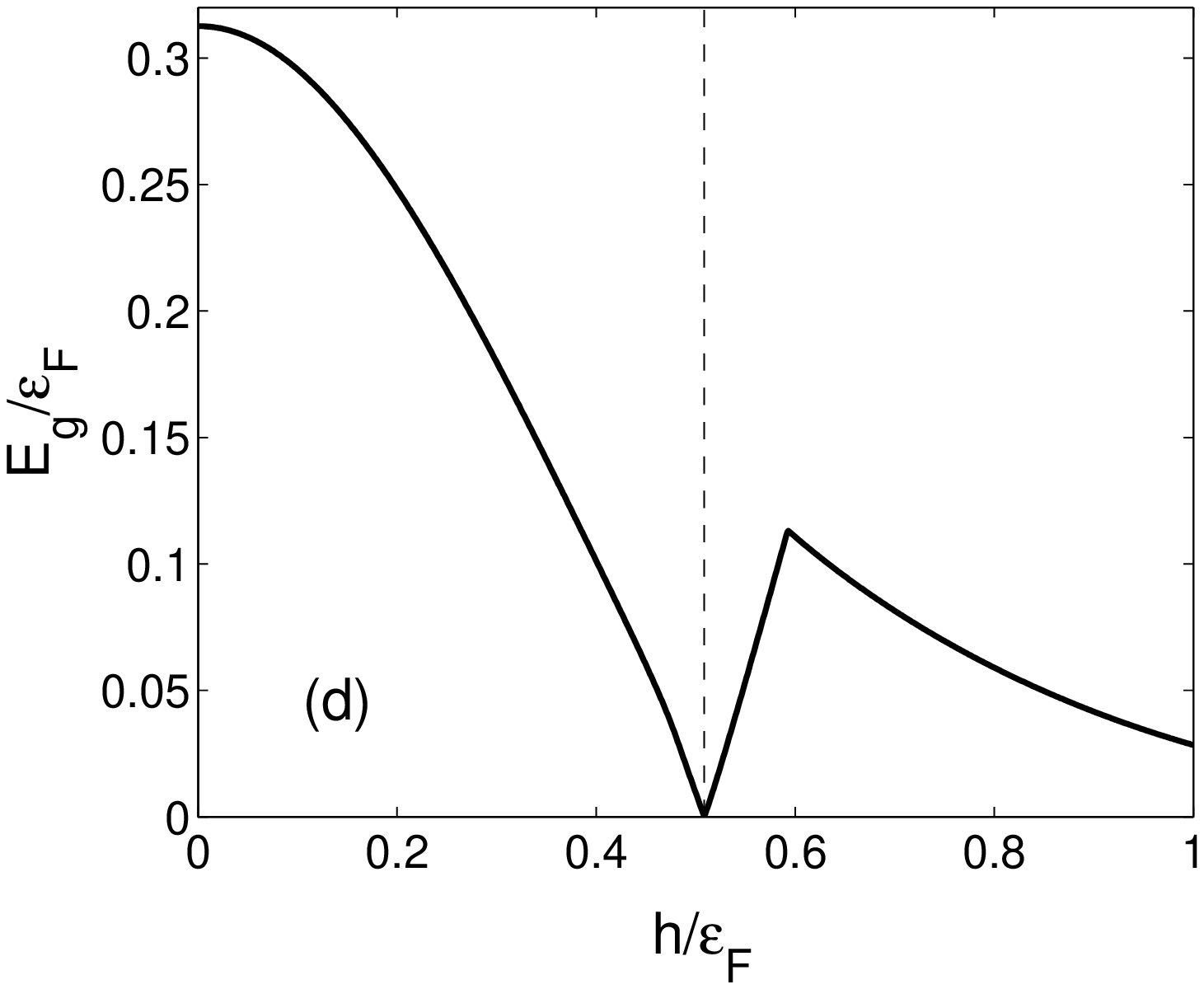}
\caption{(Color-online) Pair potential $\Delta$ (a), chemical potential $\mu$ (b), susceptibilities $\chi_{\mu\mu}$ and $\chi_{hh}$ (c), and bulk excitation gap $E_g$ (d) as functions of $h$. All quantities are properly scaled by the Fermi energy $\epsilon_{\rm F}=\pi n$. The dashed lines denote the critical ZF $h_c=(\mu^2+\Delta^2)^{1/2}$.
 \label{fig1}}
\end{center}
\end{figure*}

\begin{figure*}[!htb]
\begin{center}
\includegraphics[width=7cm]{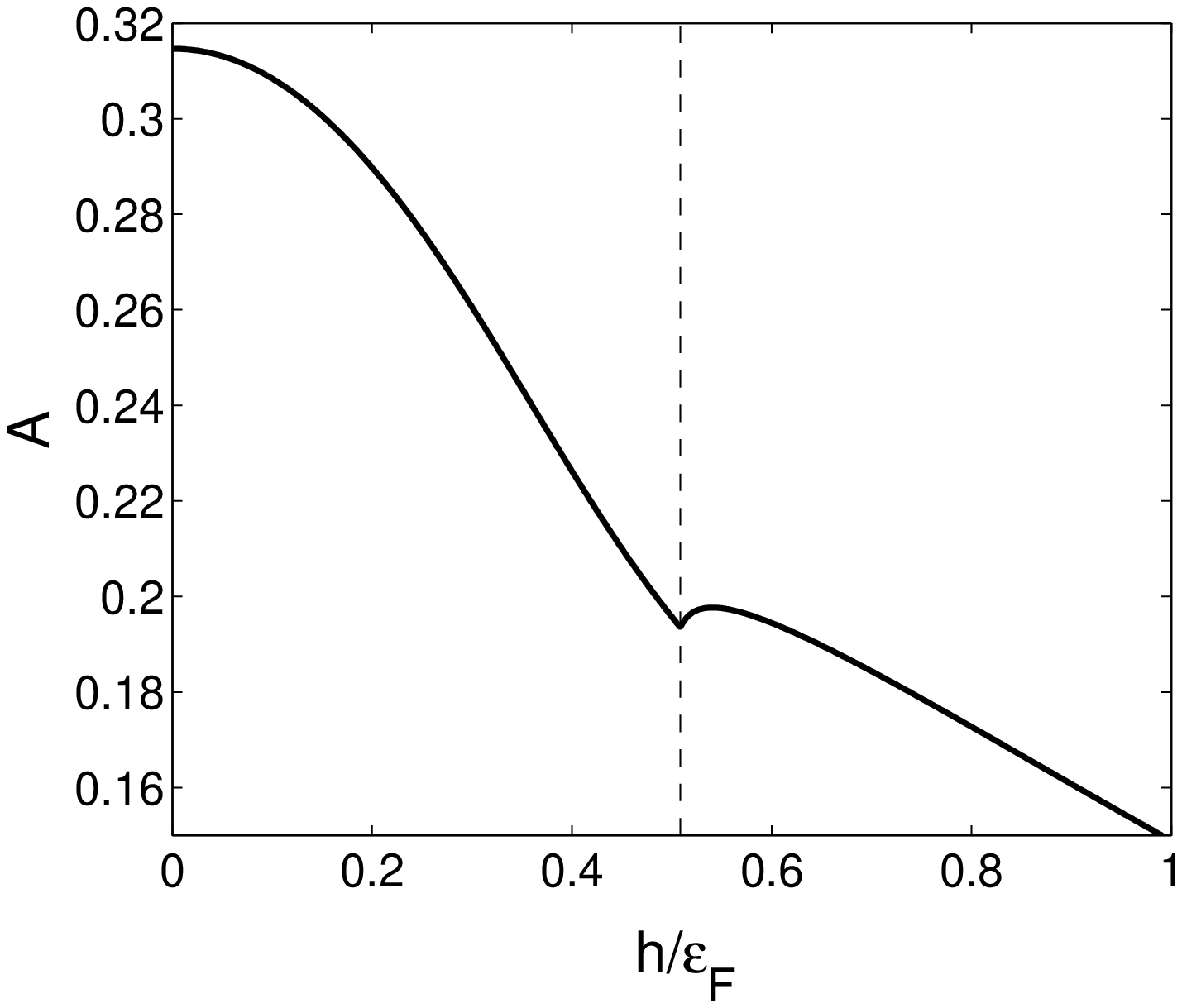}
\includegraphics[width=7cm]{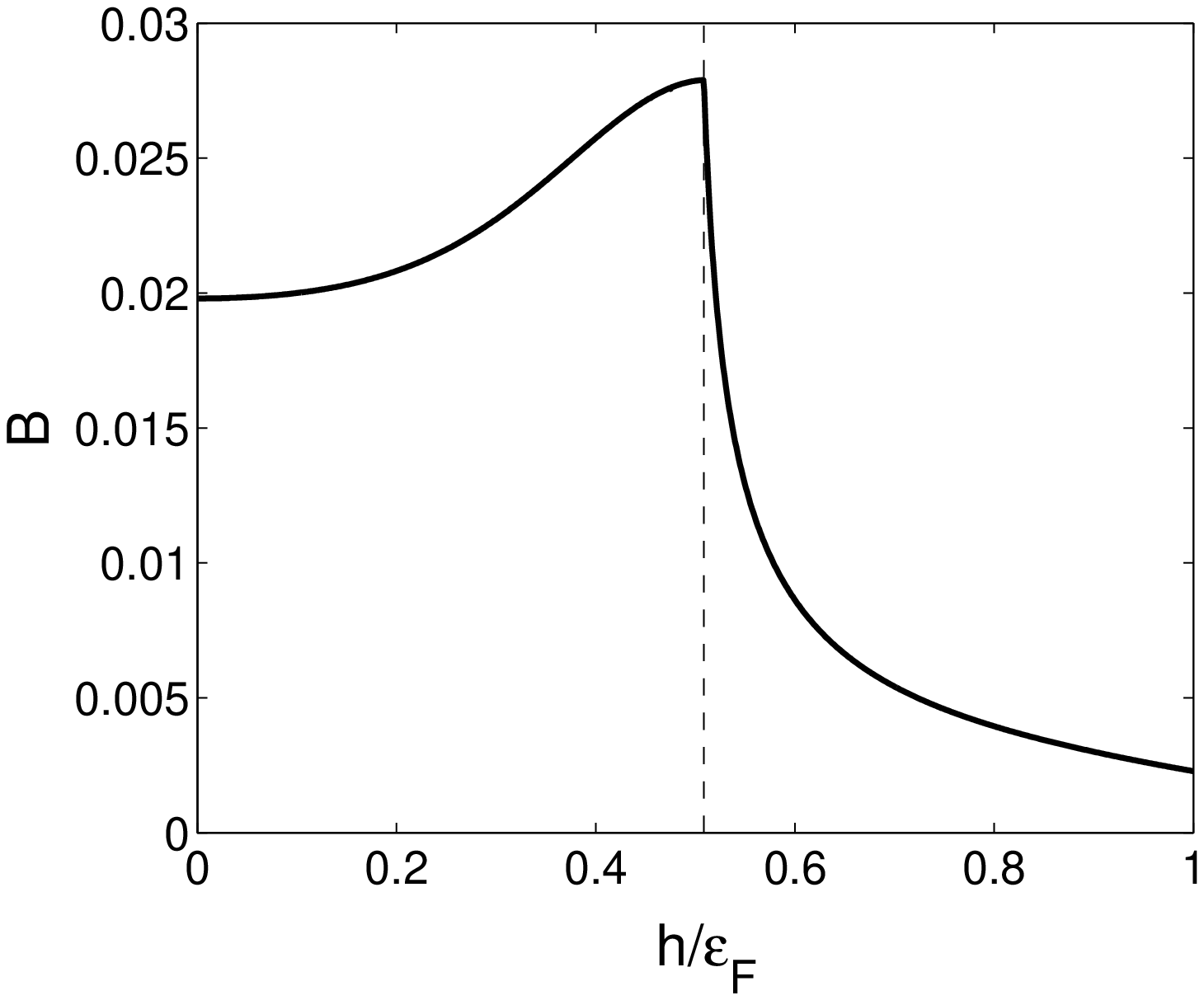}
\includegraphics[width=7cm]{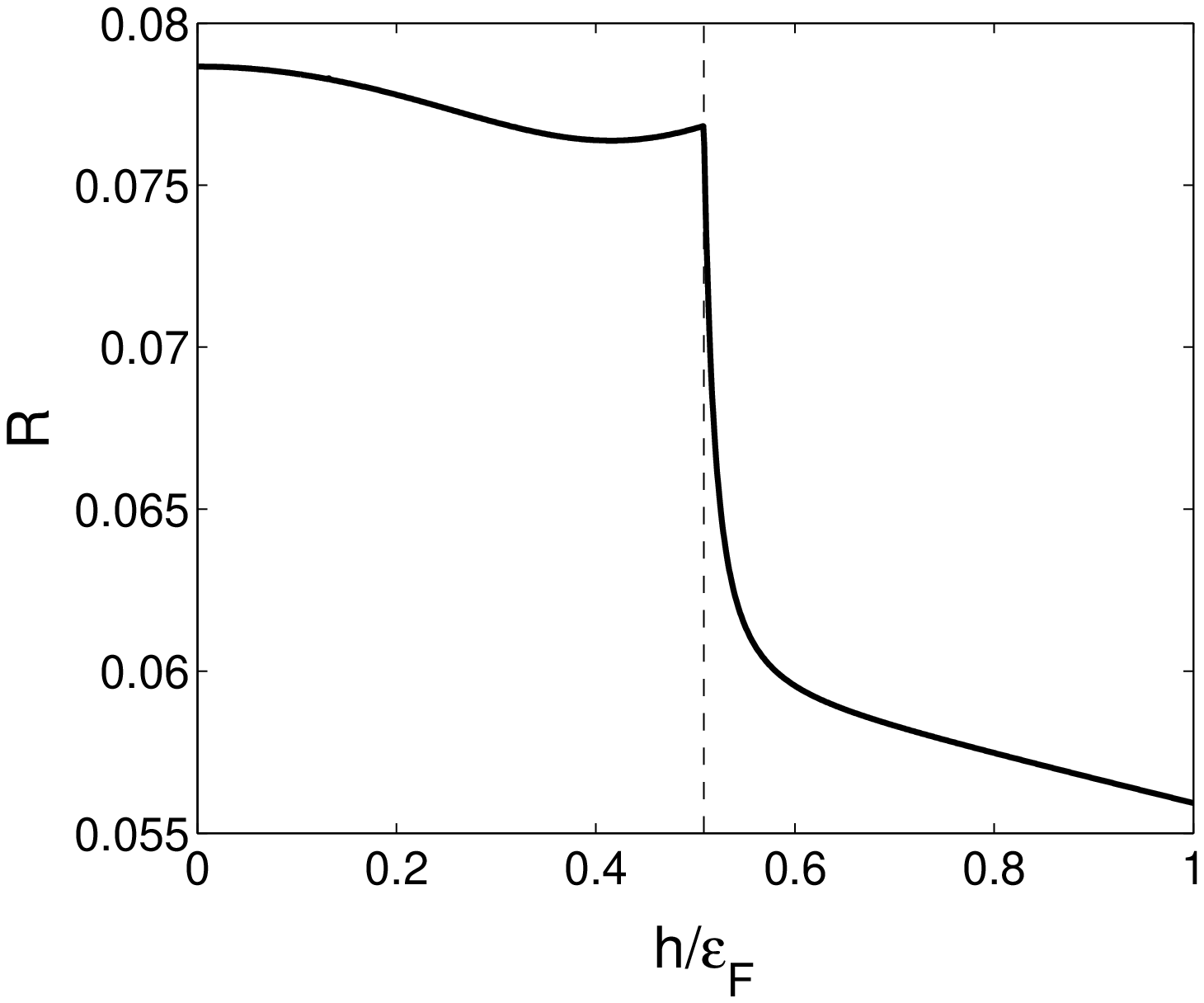}
\includegraphics[width=7cm]{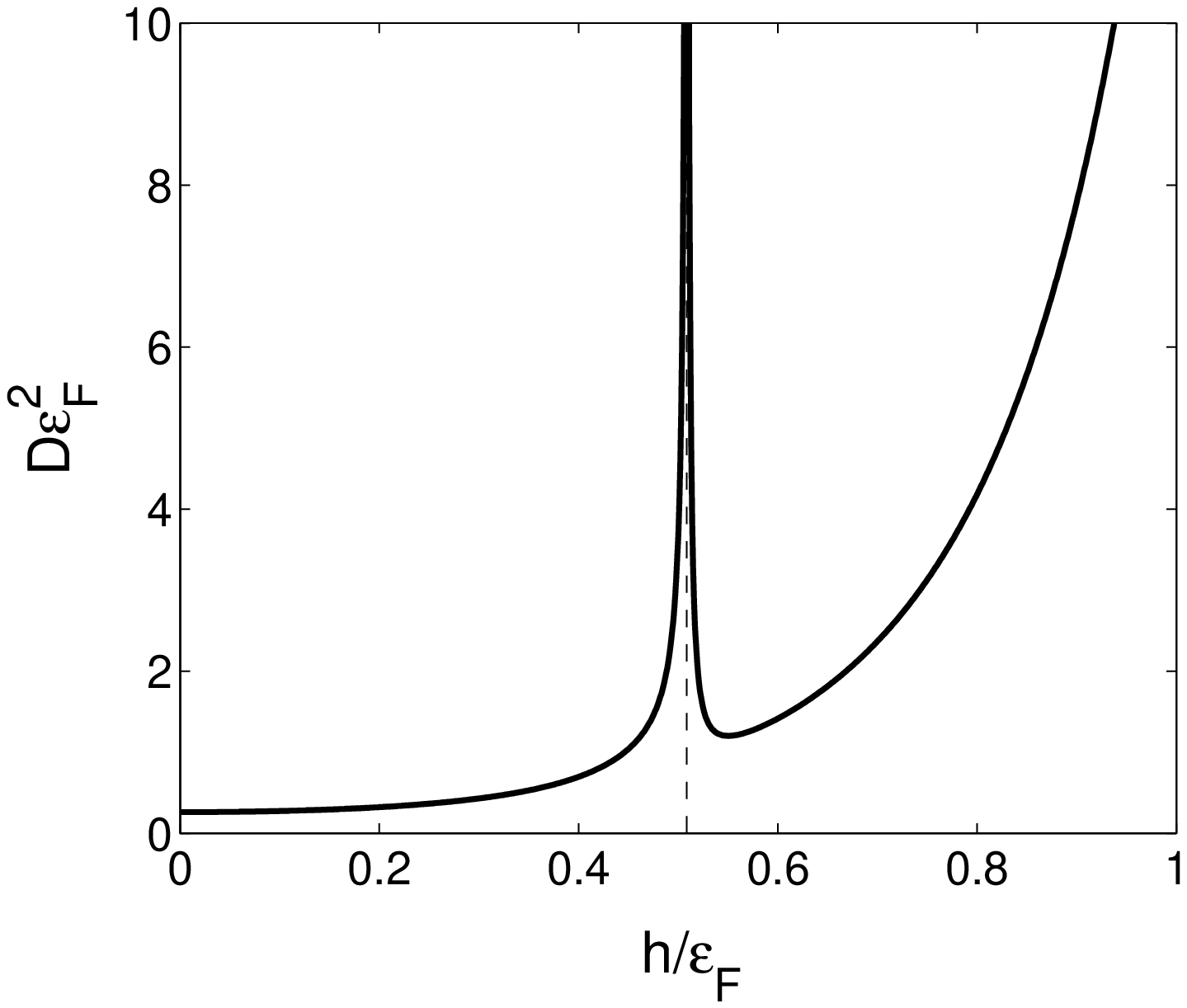}
\caption{Expansion parameters $A,B,D$, and $R$ as functions of $h$. All quantities are properly scaled by the Fermi energy
$\epsilon_{\rm F}=\pi n$.\label{fig2}}
\end{center}
\end{figure*}

\begin{figure}[!htb]
\begin{center}
\includegraphics[width=8cm]{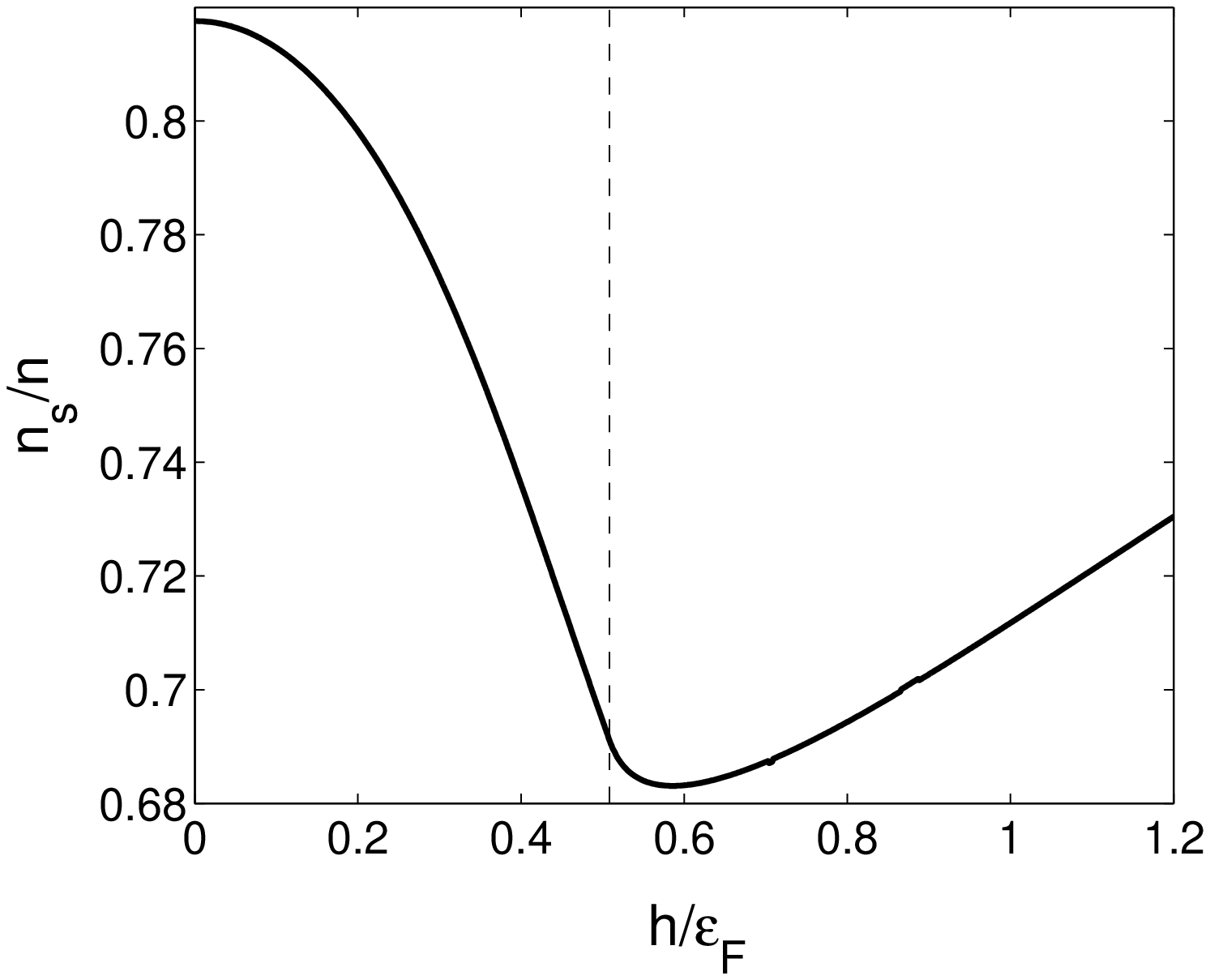}
\includegraphics[width=8cm]{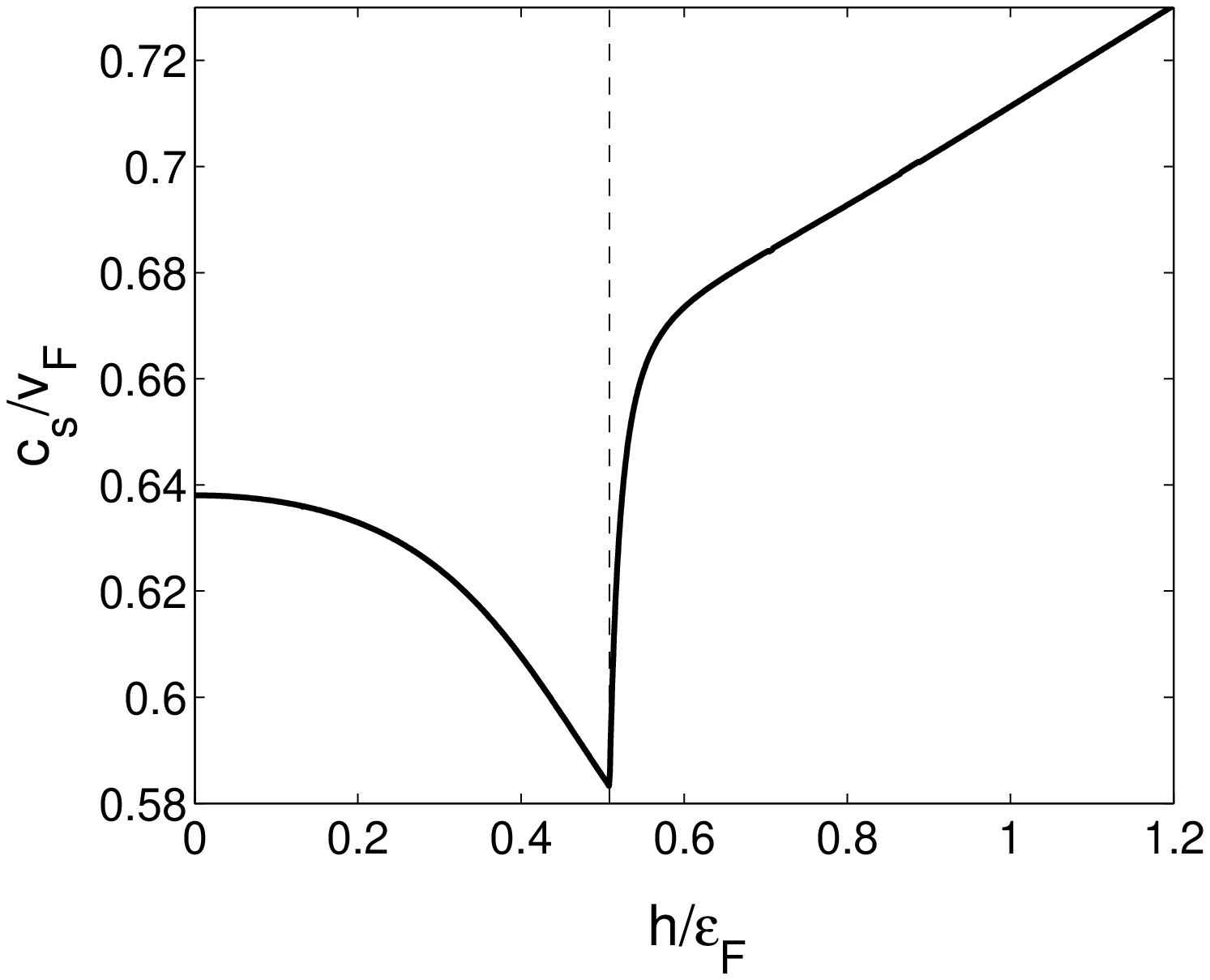}
\caption{Superfluid density $n_s$ (divided by $n$) and velocity of the Goldstone mode $c_s$
(divided the Fermi velocity $\upsilon_{\rm F}=k_{\rm F}/M$) as functions of $h$. \label{fig3}}
\end{center}
\end{figure}

For homogeneous systems, the pair potential $\Delta$ and the chemical potential $\mu$ are determined by imposing the total density
$n=k_{\rm F}^2/(2\pi)=\epsilon_{\rm F}/\pi$. The system can be characterized by two dimensionless parameters: the attractive strength
$\ln(k_{\rm F}a_{2\rm D})$ and the SOC strength $\lambda/k_{\rm F}$. Here, the 2D scattering length $a_{2\rm D}$ is defined as
$\epsilon_{\rm B}=4e^{-2\gamma}/(Ma_{2\rm D}^2)$~\cite{AS2D} with $\gamma=0.577216$ being Euler's constant. The numerical results presented in this paper are for $\ln(k_{\rm F}a_{2\rm D})=2$ and $\lambda/k_{\rm F}=0.5$. The quantum phase transition occurs at $h=h_c=(\mu^2+\Delta^2)^{1/2}\simeq0.51\epsilon_{\rm F}$. Increasing the attraction and/or SOC enhances the pairing potential and hence the critical field $h_c$, but does not lead to qualitatively different results. As shown in Fig. \ref{fig1}, the pair potential $\Delta$, although it is suppressed by the ZF, goes smoothly but never vanishes at large $h$. The chemical potential $\mu$ goes smoothly and reaches a maximum at the phase transition.  For $h>h_c$, the system is a topological superfluid~\cite{TSC01}. Figure \ref{fig1}(c) shows the susceptibilities $\chi_{\mu\mu}$ and $\chi_{hh}$. They are continuous but not smooth at the phase transition, as we expected. Figure \ref{fig1}(d) shows the bulk excitation gap $E_g=\min_{\bf k}\{E_{\bf k}^+, E_{\bf k}^-\}$. It equals the pair potential $\Delta$ only at $h=0$. Near the phase transition, it goes nonmonotonically.

For a trapped system, $h$ is fixed and the chemical potential $\mu(r)=\mu_0-V(r)$ in the local-density approximation (LDA), where $V(r)=\frac{1}{2}\omega_\perp^2r^2$ is the trap potential. In the LDA, the susceptibility $\chi_{\mu\mu}(r)$ can be obtained by the relation
\begin{equation}
\chi_{\mu\mu}(r)=-\frac{1}{\omega_\perp^2 r}\frac{dn(r)}{dr}.
\end{equation}
Therefore, the quantum phase transition can be identified by analyzing the density profile $n(r)$.

\emph{Superfluid density and collective modes -- } To study the behavior of the superfluid density $n_s$ and the collective modes across the quantum phase transition, we consider the Gaussian-fluctuation part  ${\cal S}_{\text{eff}}^{(2)}[\phi,\phi^\dagger]$. It can be written in a bilinear form
\begin{equation}
{\cal S}_{\text{eff}}^{(2)}=\frac{1}{2}\sum_Q \Lambda^\dagger(Q) {\bf M}(Q) \Lambda(Q),
\end{equation}
where $Q=(i\nu_n,{\bf q})$ with $\nu_n$ being the boson Matsubara frequency, $\Lambda(Q)=[\phi(Q), \phi^\dagger(-Q)]^{\rm T}$, and the $2\times2$ matrix ${\bf M}(Q)$ is the inverse of the collective-mode propagator. The matrix elements of ${\bf M}(Q)$ are constructed by using the fermion propagator ${\cal G}(K)$. We have
\begin{eqnarray}
{\bf M}_{11}(Q)&=&{\bf M}_{22}(-Q)\nonumber\\
&=&\frac{1}{U}+\frac{1}{2}\sum_K\text{Tr}\left[{\cal G}_{11}(K+Q){\cal G}_{22}(K)\right],\nonumber\\
{\bf M}_{12}(Q)&=&\frac{1}{2}\sum_K\text{Tr}\left[{\cal G}_{12}(K+Q){\cal G}_{12}(K)\right],\nonumber\\
{\bf M}_{21}(Q)&=&\frac{1}{2}\sum_K\text{Tr}\left[{\cal G}_{21}(K+Q){\cal G}_{21}(K)\right].
\end{eqnarray}
Taking the analytical continuation $i\nu_n\rightarrow\omega+i0^+$, the dispersions $\omega({\bf q})$ of the collective modes are
determined by the equation $\det{{\bf M}[\omega({\bf q}), {\bf q}]}=0$.

We can decompose ${\bf M}_{11}(\omega,{\bf q})$ as ${\bf M}_{11}(\omega,{\bf q})={\bf M}_{11}^+(\omega,{\bf q})+{\bf M}_{11}^-(\omega,{\bf q})$, where ${\bf M}_{11}^+(\omega,{\bf q})$ and ${\bf M}_{11}^-(\omega,{\bf q})$ are even and odd functions of $\omega$, respectively. Meanwhile 
${\bf M}_{12}(\omega,{\bf q})$ and ${\bf M}_{21}(\omega,{\bf q})$ are even functions of $\omega$ and can be expressed as
${\bf M}_{12}(\omega,{\bf q})={\bf M}_{21}^*(\omega,{\bf q})={\bf M}_{12}^+(\omega,{\bf q})+i{\bf M}_{12}^-(\omega,{\bf q})$. The term 
${\bf M}_{12}^-(\omega,{\bf q})\propto h\lambda^2$ vanishes when $h$ or $\lambda$ is zero. Then we decompose the complex field $\phi(x)$ into its amplitude mode $\rho(x)$ and phase mode $\theta(x)$, $\phi(x)=\rho(x)+i\Delta\theta(x)$. The effective action ${\cal S}_{\text{eff}}^{(2)}$ then takes the form
\begin{equation}
{\cal S}_{\text{eff}}^{(2)}=\frac{1}{2}\sum_Q\left(\begin{array}{cc} \rho(-Q)&\theta(-Q)\end{array}\right){\bf N}(Q)\left(\begin{array}{c} \rho(Q)\\
\theta(Q)\end{array}\right),
\end{equation}
where the matrix ${\bf N}(Q)$ reads ${\bf N}_{11}(Q)=2({\bf M}_{11}^++{\bf M}_{12}^+)$, ${\bf N}_{22}(Q)=2\Delta^2({\bf M}_{11}^+-{\bf M}_{12}^+)$,
${\bf N}_{12}(Q)=2i\Delta({\bf M}_{11}^--i{\bf M}_{12}^-)$, and ${\bf N}_{21}(Q)=-2i\Delta({\bf M}_{11}^-+i{\bf M}_{12}^-)$. Since 
${\bf M}_{11}^-(0,{\bf q})=0$ and ${\bf M}_{12}^-(\omega, {\bf 0})=0$, the amplitude and phase modes decouple completely at $(\omega,{\bf q})
=(0,{\bf 0})$. At the saddle point we have precisely ${\bf M}_{11}^+(0,{\bf 0})={\bf M}_{12}^+(0,{\bf 0})$. Therefore, the phase mode at ${\bf q}=0$ is gapless; that is, the Goldstone sound mode or the Anderson-Bogoliubov mode for neutral Fermi superfluids.

To study the low-energy behavior of the collective modes, we make a small ${\bf q}$ and $\omega$ expansion of ${\bf N}(Q)$ at zero temperature. In general, the expansion takes the form ${\bf N}_{11}= A+C{\bf q}^2-D\omega^2+\cdots$, ${\bf N}_{22}= J{\bf q}^2-R\omega^2+\cdots$, and
${\bf N}_{12}={\bf N}_{21}^*=-iB\omega+\cdots$. The term ${\bf M}_{12}^-(\omega,{\bf q})$ does not contribute in this expansion. The explicit forms of the expansion parameters are given by \cite{detail}
\begin{eqnarray}
&&A=\frac{1}{2}\sum_{\alpha=\pm}\sum_{\bf k}\left[\frac{\Delta^2}{(E_{\bf k}^\alpha)^3}\left(1+\alpha\frac{h^2}{\zeta_{\bf k}}\right)^2
+\alpha\frac{h^4\Delta^2}{E_{\bf k}^\alpha\zeta_{\bf k}^3}\right],\nonumber\\
&&B=\frac{\Delta}{4}\sum_{\alpha\pm}\sum_{\bf k}\Bigg[\frac{\xi_{\bf k}}{(E_{\bf k}^\alpha)^3}
\left(1+\alpha\frac{\lambda^2{\bf k}^2}{\zeta_{\bf k}}-\frac{h^2E_{\bf k}^2}{\zeta_{\bf k}^2}\right)\nonumber\\
&&\ \ \ \ \ \ \ \ +\ \frac{4\xi_{\bf k}}{(E_{\bf k}^++E_{\bf k}^-)^2}\frac{h^2}{\zeta_{\bf k}^2}
\frac{E_{\bf k}^2+\alpha\zeta_{\bf k}}{E_{\bf k}^\alpha}\Bigg],\nonumber\\
&&D=\frac{1}{8}\sum_{\alpha=\pm}\sum_{\bf k}\left[\frac{(E_{\bf k}^\alpha)^2-\Delta^2}{(E_{\bf k}^\alpha)^5}
\frac{\lambda^2{\bf k}^2\xi_{\bf k}^2}{\zeta_{\bf k}^2}+\frac{\Delta^2}{(E_{\bf k}^\alpha)^5}
\frac{\lambda^2{\bf k}^2h^2}{\zeta_{\bf k}^2}\right]\nonumber\\
&&\ \ \ \ \ \ \ \ +\sum_{\bf k}\frac{1}{(E_{\bf k}^++E_{\bf k}^-)^3}\frac{h^2\xi_{\bf k}^2}{\zeta_{\bf k}^2}
\left(1+\frac{E_{\bf k}^2-\eta_{\bf k}^2}{E_{\bf k}^+E_{\bf k}^-}\right),\nonumber\\
&&R=\sum_{\bf k}\frac{\Delta^2}{(E_{\bf k}^++E_{\bf k}^-)^3}\frac{h^2E_{\bf k}^2}{\zeta_{\bf k}^2}
\left(1+\frac{E_{\bf k}^2-\eta_{\bf k}^2}{E_{\bf k}^+E_{\bf k}^-}+\frac{2\lambda^2{\bf k}^2\Delta^2}
{E_{\bf k}^+E_{\bf k}^-E_{\bf k}^2}\right)\nonumber\\
&&\ \ \ \ \ \ \ \ +\ \frac{1}{8}\sum_{\alpha=\pm}\sum_{\bf k}\frac{\Delta^2}{(E_{\bf k}^\alpha)^3}
\frac{\lambda^2{\bf k}^2\xi_{\bf k}^2}{\zeta_{\bf k}^2},\nonumber\\
&&J=\frac{n}{4M}-\frac{1}{4M}\sum_{\alpha=\pm}\sum_{\bf k}\frac{\lambda^2}{2E_{\bf k}^\alpha}
\Bigg[\left(1-\frac{\lambda^2{\bf k}^2\xi_{\bf k}^2}{2\zeta_{\bf k}^2}\right)\nonumber\\
&&\ \ \ \ \ \ \ +\ \alpha\left(1+\frac{h^2E_{\bf k}^2}{\zeta_{\bf k}^2}
+\frac{\lambda^2{\bf k}^2h^2\Delta^2}{\zeta_{\bf k}^2E_{\bf k}^2}\right)\frac{E_{\bf k}^2}{2\zeta_{\bf k}}\Bigg].
\end{eqnarray}
The parameter $A$ equals the quantity $\partial^2\Omega/\partial\Delta^2$ in (5) at the saddle point. The phase stiffness $J$ is related to the superfluid density $n_s$ by $J=n_s/(4M)$ ($M=1$ in our units). $n_s$ can also be obtained from its standard definition \cite{NS01}. When the superfluid moves with a uniform velocity $\mbox{\boldmath{$\upsilon$}}_s$, the pair field transforms as 
$\Phi\rightarrow \Phi e^{2iM\mbox{\boldmath{$\upsilon$}}_s\cdot {\bf r}}$. The superfluid density $n_s$ is defined as the response of the thermodynamic potential $\Omega$ to an infinitesimal velocity $\mbox{\boldmath{$\upsilon$}}_s$; that is,
$\Omega(\mbox{\boldmath{$\upsilon$}}_s)=\Omega({\bf 0})+\frac{1}{2}n_s\mbox{\boldmath{$\upsilon$}}_s^2+O(\mbox{\boldmath{$\upsilon$}}_s^4)$.

Analyzing the infrared behavior of the momentum integrals, the analyticities of the expansion parameters across the phase transition can be summarized as follows: (1) The phase stiffness $J$ and hence the superfluid density $n_s$ is smooth; (2) The parameters $A,B,R$ are continuous but not smooth; (3) $D$ is divergent. The numerical results for these expansion parameters and the sound velocity
\begin{equation}
c_s=\sqrt{\frac{J}{R+B^2/A}}
\end{equation}
in the homogeneous system are shown in Figs. \ref{fig2} and \ref{fig3}. Note that the superfluid density does not equal the total density $n$ even at $h=0$ due to the lack of Galilean invariance in the presence of SOC~\cite{SOC-Ns}. Due to the nonanalyticities of $A,B$ and $R$, the sound velocity $c_s$ also shows nonanalyticity at the phase transition. Moreover, we find that $n_s$ and $c_s$ behave oppositely in the normal and the topological superfluid phases. They are suppressed by the ZF in the normal superfluid phase, but get enhanced by the ZF in the topological superfluid phase. This is quite unusual since we generally expect that the superfluidity should be suppressed by the ZF. On the other hand, the divergence of $D$ indicates that the amplitude or Higgs mode becomes a soft mode around the phase transition.

\emph{Analytical results for large Zeeman field -- } To understand the unusual behaviors of $n_s$ and $c_s$ in the topological superfluid phase, it is useful to reexpress the mean-field theory in the helicity representation~\cite{TSC02}. The helicity basis $(\psi_+,\psi_-)^{\rm T}$ is related to the ordinary basis $(\psi_\uparrow,\psi_\downarrow)^{\rm T}$ by a SU$(2)$ transformation. In the helicity basis the single-particle Hamiltonian is diagonal; that is, $H_{\rm s}=\sum_{\alpha=\pm}\sum_{\bf k}\xi_{\bf k}^\alpha \psi^\dagger_\alpha({\bf k})\psi^{\phantom{\dag}}_\alpha({\bf k})$ where $\xi_{\bf k}^\alpha=\xi_{\bf k}+\alpha\eta_{\bf k}$. Therefore, the system can be viewed as a two-band system. The ZF provides a band gap $2h$ at ${\bf k}=0$. In the presence of pairing, the mean-field approximation for $H_{\rm int}$ reads
\begin{equation}
H_{\rm int}\simeq\frac{1}{2}\sum_{\alpha,\beta=\pm}\sum_{\bf k}
\left[\Delta_{\alpha\beta}({\bf k})\psi_\alpha^\dagger({\bf k})\psi^\dagger_\beta(-{\bf k})+\rm{H.c.}\right].
\end{equation}
The new ${\bf k}$-dependent pair potentials $\Delta_{\alpha\beta}({\bf k})$ read $\Delta_{+-}({\bf k})=-\Delta_{-+}({\bf k})
=-\Delta_{\rm s}({\bf k})$ and $\Delta_{++}({\bf k})=\Delta^*_{--}({\bf k})=-\Delta_{\rm t}({\bf k})$, where the interband and the intraband pair potentials are given by $\Delta_{\rm s}({\bf k})=h\Delta/\eta_{\bf k}$ and $\Delta_{\rm t}({\bf k})=\lambda(k_x-ik_y)\Delta/\eta_{\bf k}$.
Using these new pair potentials, the quasiparticle dispersions $E_{\bf k}^\pm$ can be expressed as
\begin{eqnarray}
\label{helicity5}
E_{\bf k}^\pm=\sqrt{\left[\sqrt{\xi_{\bf k}^2+|\Delta_{\rm s}({\bf k})|^2}\pm\eta_{\bf k}\right]^2+|\Delta_{\rm t}({\bf k})|^2}.
\end{eqnarray}

For $h\gg h_c$, we find that the pair potential goes as $\Delta\simeq a/h^2$, while the chemical potential $\mu\simeq -h+b$, where $a$ and $b$ are some constants and $b\ll h$. Therefore, the upper band with dispersion $\xi_{\bf k}^+$ has a large gap and essentially plays no role in fermion pairing. The lower band $\xi_{\bf k}^-$ opens a Fermi surface at
\begin{equation}
k=\tilde{k}_{\rm F}=\sqrt{2\left[\lambda^2+\mu+\sqrt{\lambda^4+2\lambda^2\mu+h^2}\right]}.
\end{equation}
Since the pair potential $\Delta\ll h$, the total density $n$ is carried by the lower band. We have $n\simeq\sum_{\bf k}\Theta(\eta_{\bf k}-\xi_{\bf k})=\tilde{k}_{\rm F}^2/(4\pi)$ where $\Theta(x)$ is the standard step function, and hence $\tilde{k}_{\rm F}\simeq\sqrt{2}k_{\rm F}$. Then the system can be regarded as a weakly coupled $p_x+ip_y$ superfluid of spinless fermions where the pairing occurs around the Fermi surface $k=\tilde{k}_{\rm F}$. The interband pair potential $\Delta_{\rm s}({\bf k})$ can be safely dropped and we have 
$E_{\bf k}^-\simeq[(\xi_{\bf k}^-)^2+|\Delta_{\rm t}({\bf k})|^2]^{1/2}$. Near the Fermi surface, we get 
$E_{\bf k}^-\simeq[\tilde{\upsilon}_{\rm F}^2(k-\tilde{k}_{\rm F})^2+E_g^2]^{1/2}$ where the Fermi velocity 
$\tilde{\upsilon}_{\rm F}\simeq\sqrt{2}\upsilon_{\rm F}(1-\lambda^2/\eta_{\rm F})$ and the bulk excitation gap reads $E_g\simeq\Delta\lambda\tilde{k}_{\rm F}/\eta_{\rm F}$. Here we have defined $\eta_{\rm F}=(\lambda^2\tilde{k}_{\rm F}^2+h^2)^{1/2}$.

Based on the above observations, the superfluid density $n_s$ can be approximated as
\begin{equation}
n_s\simeq n\left(1-\frac{\lambda^2}{\eta_{\rm F}}\right).
\end{equation}
Therefore, for $h\rightarrow\infty$, we have $n_s\rightarrow n$. It manifests the fact that, for large $h$, the pairing occurs only in the lower band which carries nearly the total density. Meanwhile, the other expansion parameters $A,B$ and $R$ are dominated by the terms that are peaked at the Fermi surface $k=\tilde{k}_{\rm F}$. Using the same integral technique in BCS theory, we obtain $B^2/A\simeq0$ and $R\simeq1/[8\pi(1-\lambda^2/\eta_{\rm F})]$. Therefore, the sound velocity $c_s\rightarrow \upsilon_{\rm F}$ for $h\rightarrow\infty$. This result can be reexpressed as
\begin{equation}
c_s\simeq\frac{\tilde{\upsilon}_{\rm F}}{\sqrt{2}},
\end{equation}
which is just the sound velocity of weakly coupled 2D Fermi superfluids.

These analytical results show that, as the ZF is increased, the system behaves more and more like a $p_x+ip_y$ superfluid of spinless fermions. Therefore, the fermion pairing in the topological superfluid phase feels less stress than in the normal superfluid phase. This explains the unusual behaviors of $n_s$ and $c_s$ at large ZF.

\emph{Indication for $p$- and $d$-wave pairings -- } Finally, we point out that the infrared singularities which cause the nonanalyticities should also show up in other systems, such as the 2D BCS-BEC evolution with $p$- and $d$-wave pairings~\cite{PIP}. In such systems, the single-particle excitation spectrum is $E_{\bf k}=[\xi_{\bf k}^2+|\Delta({\bf k})|^2]^{1/2}$, where $\Delta({\bf k})\sim k$ for $p$-wave and $\Delta({\bf k})\sim k^2$ for $d$-wave pairings. At the quantum critical point $\mu=0$, the dispersion at low $k$ goes as $E_{\bf k}\sim k$ for $p$-wave and $E_{\bf k}\sim k^2$ for $d$-wave pairings. Therefore, we expect that the collective-mode properties in such systems also show nonanalyticities. The nonanalytical behavior of the collective modes can be measured by using Bragg spectroscopy~\cite{Bragg}.

\emph{Acknowledgments} --- The work is supported by the Helmholtz
International Center for FAIR within the framework of the LOEWE
program. XGH
also acknowledges the support from Indiana University Bloomington.


\begin{thebibliography}{99}
\bibitem{note01}      {Precisely speaking, the quantum phase transition is of $n$-th order, where $2<n\leq 3$.}
\end{thebibliography}
\end{document}